# ITER ECE front-end design, alignment and in-situ calibration

Saeid Houshmandyar,[1] W. L. Rowan,[1] J. P. Ziegel[1] and A. Ouroua[2]

[1] *Institute for Fusion Studies, The University of Texas at Austin, Austin, TX 78712, USA*
[2] *Center for Electromagnetics, The University of Texas at Austin, Austin, TX 78758, USA*

## Abstract

The electron cyclotron emission (ECE) diagnostics suite at ITER utilizes a front-end quasi-optical (QO) system whose design is fundamentally constrained by a field-stop concept. The field-stop defines the Gaussian beam variation throughout the optical system and within the plasma, thereby setting the ECE sampling volume and spatial resolution. An in-situ hot calibration source, optimized using Gaussian beam transmission criteria, provides independent and absolute electron temperature measurements. The QO system extends beyond the front-end to include the polarization splitter unit (PSU), transmission lines, and switchyard, forming an integrated optical path to the ECE instruments. Misalignment between the front-end and PSU reduces the effective field-stop size, degrading spatial resolution and measurement fidelity. The oblique ECE view, a key feature of the ITER design, enhances sensitivity to non-thermal electron populations and complements the diagnosis of neoclassical tearing modes. Integrated QO design and plasma physics understanding are essential for reliable ITER ECE measurements.

## 1. Introduction

Electron cyclotron emission (ECE) diagnostics exploit the natural radiation emitted by gyrating electrons in a magnetized plasma, providing a direct and localized measurement of electron thermal properties. Because the emission is intrinsically linked to the optically thick electron population at the EC resonant layer, ECE enables high spatial and temporal resolution measurements of the electron temperature ($T_e$) and its fluctuations.[1] These capabilities make ECE a fundamental diagnostic for investigating transport, monitoring plasma stability,[2] and supporting control-oriented applications in modern fusion devices.[3,4]

The diagnostic mission of ITER is to enable robust plasma operation, ensure machine protection, and support the scientific understanding required for future tokamak-based fusion power plants. The ITER ECE diagnostic[5,6] directly contributes to this mission by providing high-resolution, non-invasive measurements of the electron temperature profile. These measurements are essential for transport studies and model validation, real-time profile control (including heating and current drive), and equilibrium reconstruction. In addition, ECE supports real-time plasma control through feedback stabilization of instabilities such as neoclassical tearing modes (NTMs),[7] and by monitoring electron temperature gradients[8,9] relevant to transition thresholds (*e.g.*, L-H transition). ECE measurements also play an important role in machine protection by enabling the detection of rapid changes in the $T_e$-profile, such as those associated with disruptions, thereby triggering protective actions. Furthermore, ECE provides complementary information to other diagnostics, including magnetics, bolometry, and Thomson scattering, allowing cross-validation of plasma measurements.

The design and development of the ITER ECE diagnostic system are shared between the United States and India Domestic Agencies (US-DA and IN-DA). The system includes two plasma viewing geometries (radial and oblique) where each consists of a front-end quasi-optical (QO) system,[10] a grid polarizer to separate X- and O-mode emissions, low-loss transmission lines, a switchyard, and a suite of ECE instruments located in the diagnostic hall. These instruments include low- and high-frequency radiometers as well as Fourier transform spectrometers (FTS), enabling measurements over a broad frequency range of approximately 75-1000 GHz. The system is designed to operate at both the nominal ITER magnetic field (5.3 Tesla) and half-field (2.65 Tesla). A key aspect of the ITER ECE front-end design is the use of a field-stop-driven quasi-optical framework, which governs beam propagation, calibration, and spatial resolution.

This paper focuses on the design and performance of the ITER ECE front-end system. Section 2 describes the quasi-optical design of the front-end. Section 3 presents the in-situ calibration source and its optimization based on Gaussian beam analysis. Section 4 examines the impact of misalignment between the front-end and PSU on diagnostic performance. Section 5 discusses the role of the oblique viewing geometry in accessing non-thermal electron populations and in supporting NTM diagnosis. The manuscript is summarized in Section 6.

## 2. ITER ECE front-end

The layout of the ITER ECE front-end is shown in Figure 1. The front-end is housed within the second (middle) diagnostic shielding module (DSM2) located in equatorial port 9 (EP9) of the ITER vacuum vessel. It provides QO paths for both radial and oblique[11] views, corresponding to the lower (blue) and upper (green) channels in Figure 1, respectively.

For each viewing geometry, the collected emission is first reflected by an ellipsoidal mirror that focuses radiation from the plasma. A flat mirror then redirects the beam toward a window assembly (WA), which separates the in-vessel vacuum environment from the external QO transmission path. Downstream of the WA, the beam propagates through a displacement compensation unit and is subsequently limited by a field-stop, before continuing toward a polarization splitter unit (PSU). A three-dimensional rendering of the WA and field-stop housing is shown in Figure 2.

The DSM2 is subdivided into six bays, separated by neutron shielding panels to mitigate neutron fluence effects on the optical components. Each shielding panel includes precision bores designed to accommodate Gaussian beam propagation while maintaining shielding effectiveness.

Each view is equipped with an independent in-situ hot calibration source (HS), enabling absolute and independent electron temperature measurements. A shutter mirror, actuated from bay 5 of DSM2, switches the optical path between plasma emission and the calibration source during calibration cycles.

The vacuum boundary is defined at the closure plate (not shown), where the WA is located. The WA consists of two quartz windows, one of which is tilted by 2° to suppress interference effects. After transmission through the WA, the beam passes through a silicon carbide field-stop and the displacement compensation unit before reaching the PSU, where X- and O-mode polarizations are separated and coupled into their respective waveguides for transmission to the ECE instruments.

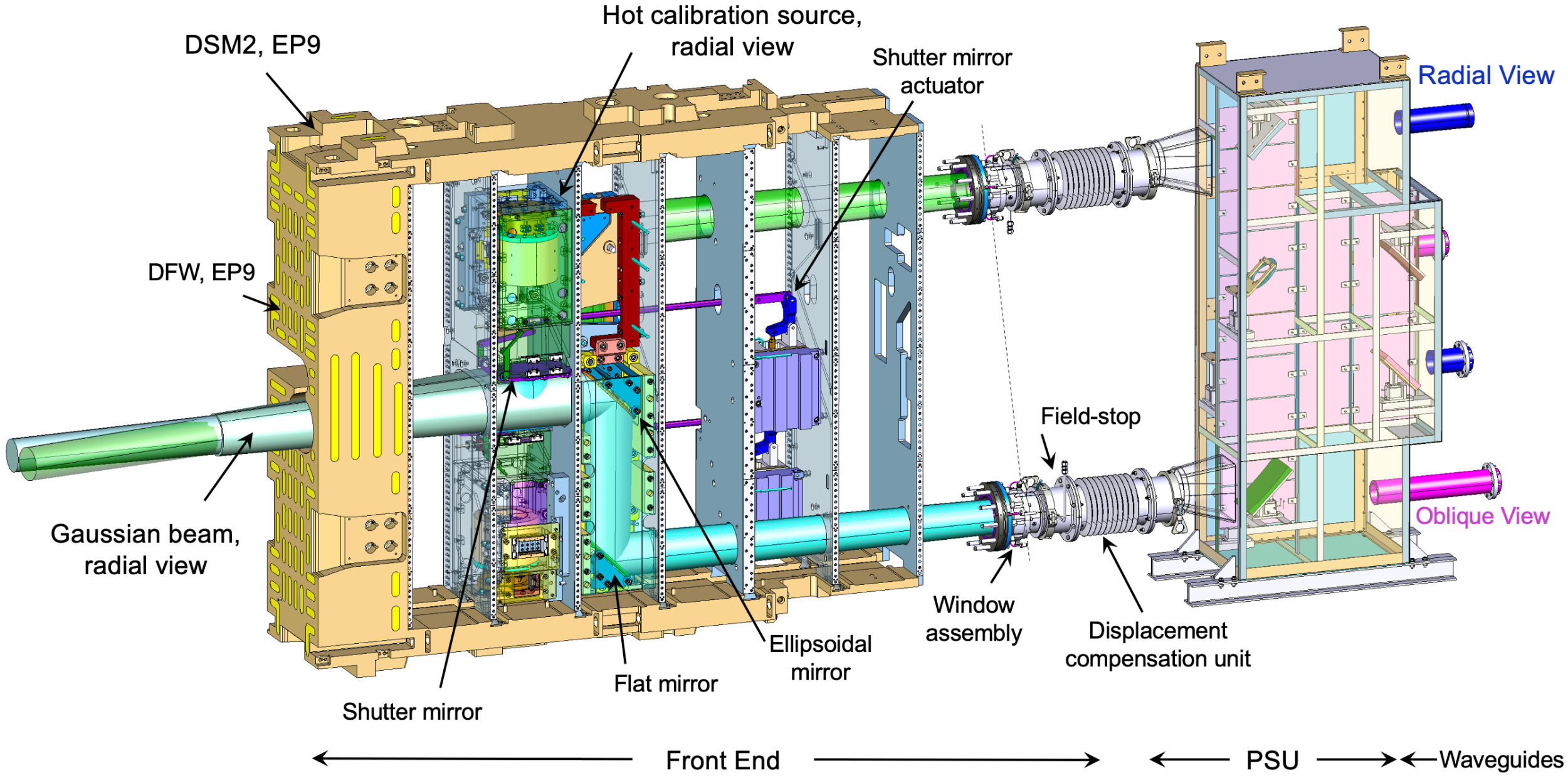

**Figure 1.** ITER-ECE front end located in diagnostic shielding module (DSM2) in equatorial port 9 (EP9). The radial (shown in blue) and oblique (shown in green) views are protruding from the diagnostic front wall (DFW) on the left side of the figure. Only the radial view components are labeled.

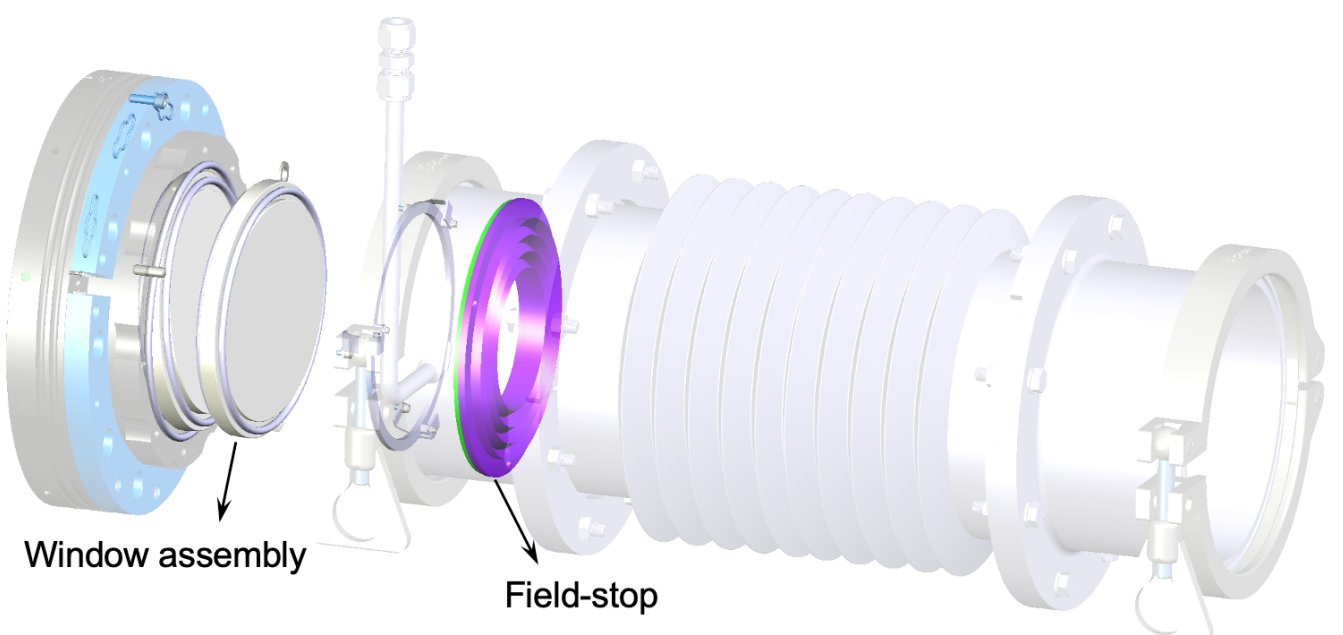

**Figure 2.** Three-dimensional rendering of the ITER ECE window assembly and field-stop housing integrated within DSM2.

The QO path within the DSM is maintained under vacuum, thereby eliminating concerns associated with molecular absorption within this region. In contrast, the transmission line downstream of the WA operates outside the vacuum environment and is therefore susceptible to atmospheric absorption effects. A primary concern is the presence of moisture, which can significantly increase attenuation of the ECE signal. As shown in Figure 3.*a*, increasing humidity not only enhances discrete molecular absorption lines,[12] but also raises the continuum absorption level, leading to overall degradation of signal transmission. To mitigate these effects, the transmission line should be purged with dry air in order to minimize moisture content and reduce attenuation. Figure 3.*a* also shows attenuation for a nitrogen-purged transmission line. While nitrogen purge effectively suppresses moisture-related absorption, its use introduces operational risks, including potential asphyxiation hazards, and must therefore be carefully managed within applicable safety constraints.

Nevertheless, even small amounts of residual moisture can significantly degrade ECE transmission over long propagation distances. Figure 3.*b* shows the calculated blackbody emission from a HS operating at 700 °C, evaluated both at the source location and after propagation through 1 m and 30 m transmission lines containing air with 5% moisture. The transmission line temperature is assumed to be at the room temperature. The results demonstrate that moisture severely attenuates the transmitted emission not only at the water absorption lines, but also through overall degradation of the high-frequency portion of the spectrum approaching 1 THz, which is particularly important for the FTS measurements.

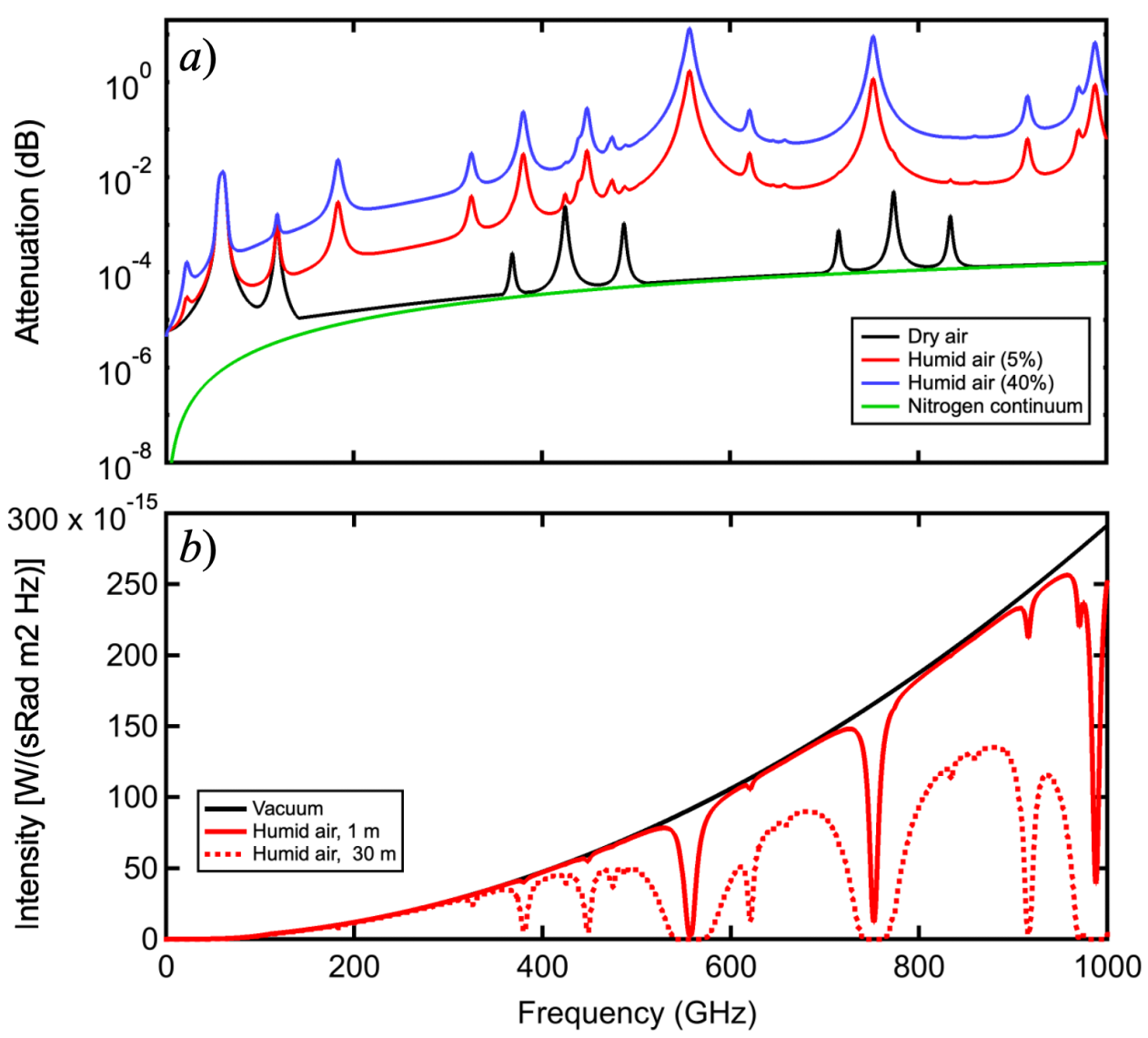

**Figure 3.** *a*) Atmospheric attenuation across the ITER ECE frequency range for dry air, humid air (5% and 40% moisture), and nitrogen purge conditions. *b*) Calculated blackbody emission from the hot calibration source at 700 °C after transmission through 1 m and 30 m atmospheric paths containing 5% moisture, with the transmission line at 20 °C.

## 3. Gaussian beam analysis

The QO design of the ITER ECE system is guided by several competing requirements. On one hand, the system must minimize the ECE sampling volume within the plasma and maintain a relatively uniform spatial resolution across the minor radius. On the other hand, it must maximize the collection and transmission of ECE power to the diagnostic instruments for both radial and oblique viewing geometries. Achieving high transmission efficiency generally favors larger optical components, including mirrors and the emissive surface of the HS, as well as WA's diameter. However, the available space within DSM2 imposes strict constraints on component dimensions. As a result, the sizing of optical elements represents a compromise between maximizing signal throughput and satisfying geometric and mechanical limitations. In particular, because the ITER ECE system is designed to provide independent absolute calibration, the diameter of the HS emissive surface sets a lower bound on the minimum frequency at which calibrated ECE measurements can be obtained.

In addition to the HS emissive surface, the ECE sampling volume is also determined by the field-stop size and the properties of the first ellipsoidal mirror, including its focal length ($f = 2100$ mm) and its relative position with respect to the plasma edge and the field-stop. The field-stop acts as a spatial filter on the collected emission, constraining the beam size and directly influencing the sampling volume within the plasma. Unlike a conventional aperture used only for beam limitation, the field-stop in the ITER ECE system defines the Gaussian beam waist and therefore governs the beam variation along the QO path. As such, it serves as the central element linking the optical design to the diagnostic spatial resolution and effectively defines the viewing geometry of the system.

In a QO system incorporating a field-stop, the Gaussian beam waist is constrained such that its radius matches the field-stop size radius that location. The propagation of the microwave beam along the QO path can be described by solutions to Maxwell's equations,[13] which under the paraxial approximation, reduce to Gaussian beam optics (Figure 4). For a beam with a minimum waist $w_0$, the beam radius $w(z)$ and wavefront curvature $R(z)$ as functions of the propagation distance $z$ from the waist are given by:

$$w(z) = w_0\sqrt{1 + \left(\frac{\lambda z}{\pi w_0^2}\right)^2} \quad \text{, and} \tag{1}$$

$$R(z) = z\left[1 + \left(\frac{\pi w_0^2}{\lambda z}\right)^2\right]. \tag{2}$$

As the Gaussian beam is reflected from a concave mirror, the curvature of the incoming wavefront, $R_{in}$, is transformed into an outgoing curvature, $R_{out}$, according to Eq. (3), where $f$ denotes the focal length of the mirror. Note that the flat mirrors do not change the Gaussian beam widths.

$$\frac{1}{R_{in}} + \frac{1}{R_{out}} = \frac{1}{f}\,. \tag{3}$$

**Figure 4.** Irradiance profile of fundamental Gaussian mode.

Figure 5 shows the Gaussian beam waist as a function of distance along the ITER ECE QO path, referenced to the field-stop location, for several frequencies spanning the ITER ECE operating range from 75 GHz to 1 THz. The field-stop radius was selected to be 41.75 mm. This value was determined through an iterative optimization process involving both the field-stop size and the focal length of the first ellipsoidal mirror, with the objective of achieving a relatively uniform beam width across the plasma minor radius. As shown in Figure 5, the beam variation differs for each frequency along the optical path; however, all beam waists coincide at the field-stop location, where the beam waist matches the field-stop radius. Downstream of the field-stop, the beam propagates through the PSU, which effectively operates as a Gaussian beam telescope. As a result, the beam undergoes a demagnification of approximately 1.5:1 prior to coupling into the waveguides. The relative positions of the optical components are based on the ITER ECE front-end layout shown in Figure 1.

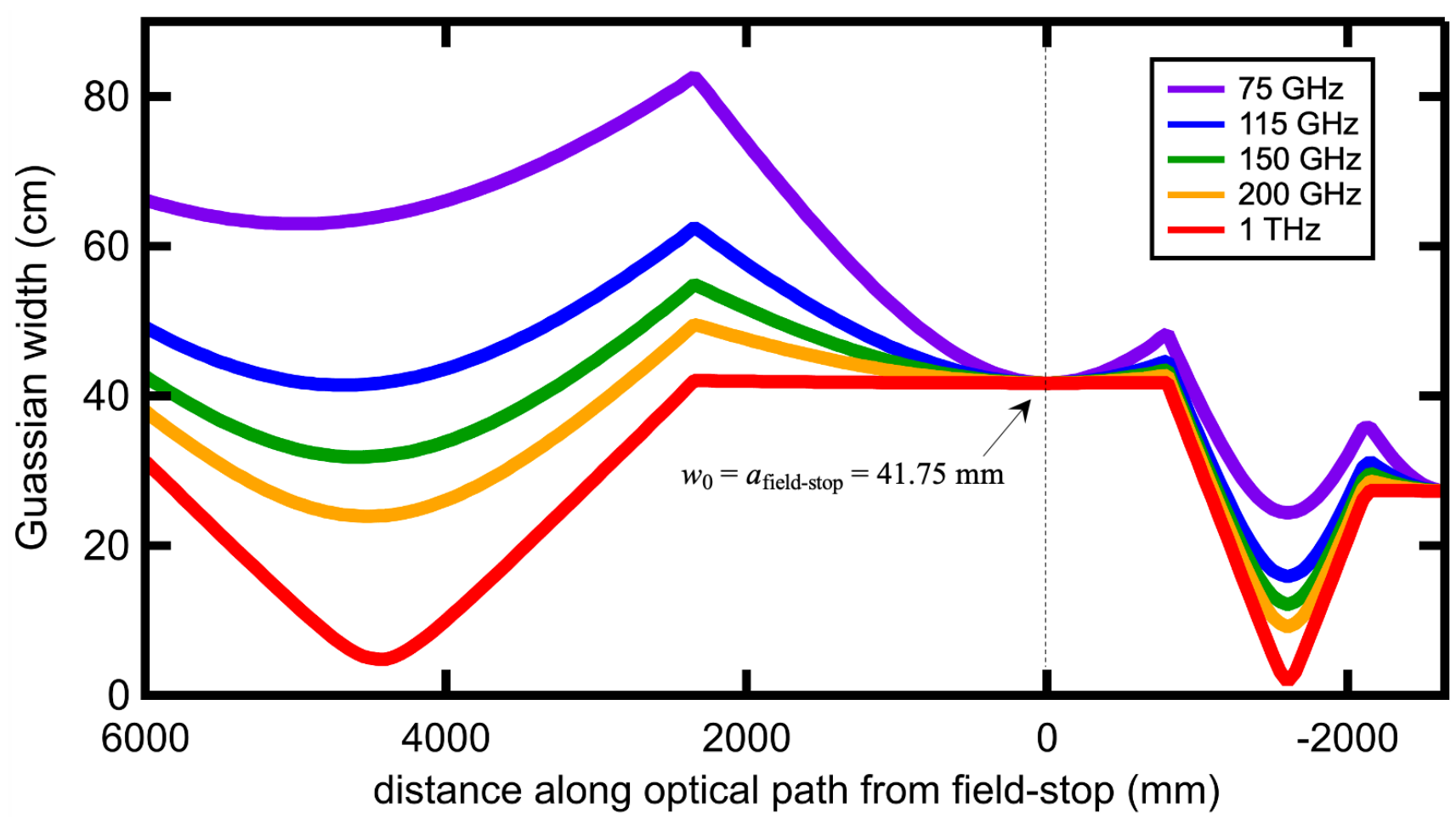

**Figure 5.** Gaussian beam width along ITER ECE quasi-optical path, referenced to the field-stop is located at 0. Positive distances correspond to the front-end, while negative distances correspond to the PSU side; the traces terminate at the waveguide coupling locations.

The QO design discussed so far assumes ideal optical elements and does not account for the finite dimensions of the mirrors or the finite emissive surface of the HS. In practice, however, the finite size of optical components introduces Gaussian beam truncation, causing the standard QO approximation to break down. Beam truncation modifies both the transmitted power and the downstream beam profile, thereby affecting overall system performance. Since the Gaussian beam size increases at lower ECE frequencies, minimizing truncation requires sufficiently large optics and HS emissive surface sizes. In reality, the available space within DSM2 imposes strict geometric constraints on the achievable mirror and HS dimensions. Consequently, the diameter of the HS emissive surface defines the minimum calibrated ECE frequency for which efficient power transmission through the QO system can be maintained. The fractional power transmitted through a circular aperture of radius $a$ (diameter $d = 2a$) can be expressed as $P/P_0$ = 1-exp($-2a^2/w^2$),[14,15] where $P$ is the transmitted power, $P_0$ is the incident power, and $w$ is the Gaussian beam radius at the aperture location. The transmitted fractional power reaches approximately 98.88% when the aperture diameter satisfies $d = 3w$.[16] For this reason, the 3$w$ criterion is adopted as the primary guideline for sizing optical elements throughout the ITER ECE QO system. Since the beam radius is largest at the lowest operating frequencies, the optical design must satisfy the 3$w$ condition at the minimum calibrated ECE frequency to ensure efficient transmission across the full ITER ECE frequency range.

The minimum calibrated ECE frequency is determined by the diameter of the emissive surface of the in-situ HS. This is established by evaluating the 3$w$ beam diameter at the HS location over the full ITER ECE frequency range, as shown in Figure 6. The intersection between the calculated 3$w$ curve and the physical diameter of the HS emissive surface defines the minimum frequency at which the 3$w$ transmission criterion can still be satisfied. For the present ITER ECE front-end design, this frequency is found to be approximately 115 GHz.

In practice, however, the available space within DSM2 imposes significant constraints on the dimensions of the optical components. Among these components, the HS represents one of the most space-demanding elements due not only to the size of its emissive surface, but also to the additional enclosure volume associated with the thermal, mechanical, and electrical subsystems needed to support prolonged calibration operation within the ITER in-vessel. These engineering constraints has limited the achievable diameter of the HS emissive surface to 150 mm. Consequently, the 150 mm diameter establishes the minimum calibrated ECE frequency for the ITER ECE system, based on the relative location of the HS with respect to the field-stop within the DSM2 layout.

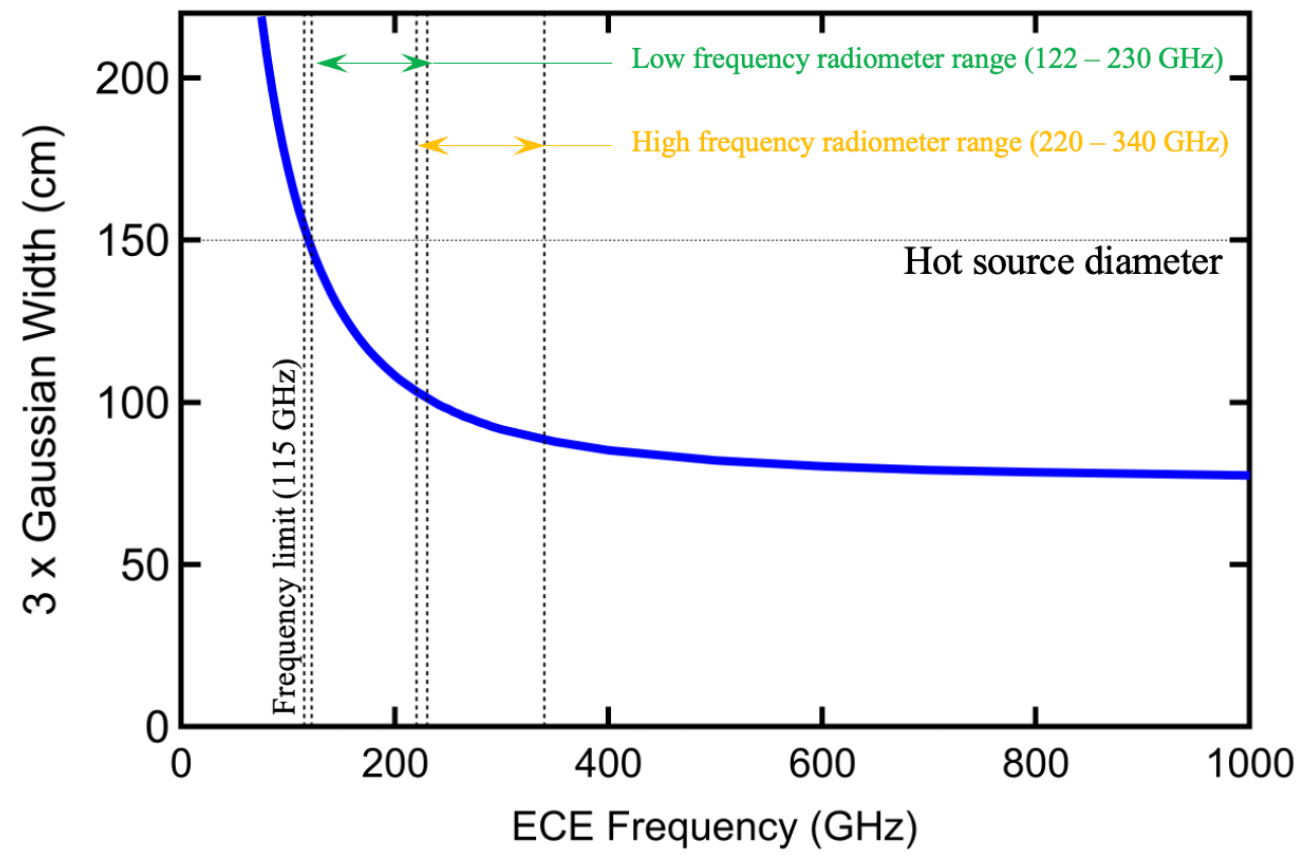

**Figure 6.** Calculated 3$w$ beam diameter at the hot calibration source location as a function of ECE frequency. The intersection with the 150 mm HS emissive surface defines the minimum calibrated ECE frequency of 155 GHz.

### 3.1. Physics design of the in-situ hot calibration source

Under the blackbody approximation, the electron cyclotron emission from an optically thick plasma is proportional to the electron temperature, $T_e$. The emitted intensity can therefore be expressed as $I = CT_e$, where $I$ is the measured emission intensity and $C$ is a proportionality constant that depends on the optical transmission and instrument response. The emitted EC radiation propagates through the ITER ECE quasi-optical system, including the front-end optics, PSU, switchyard, and waveguides, before reaching the ECE instruments located in the diagnostic hall. These instruments, including ECE radiometers and Fourier transform spectrometers (FTS), ultimately produce output signals in arbitrary units. Therefore, an absolute calibration procedure is required to relate the measured signal levels to the plasma electron temperature.

The in-situ HS provides this calibration capability by replacing the plasma emission with a controlled blackbody-like radiation source located within the same QO path. By measuring the instrument response for multiple HS temperatures, calibration coefficients can be determined as functions of frequency, thereby enabling absolute electron temperature measurements. This constitutes the fundamental physics basis for incorporating in-situ hot calibration sources into the ITER ECE diagnostic system.

The design of the HS is intended to produce radiation that closely approximates blackbody emission. The spectral radiance of a blackbody at frequency $f$ and absolute temperature $T$ is given by

$$I(f,T) = \varepsilon \frac{2hf^3}{c^2} \frac{1}{e^{\left(\frac{hf}{k_B T}\right)} - 1} . \tag{4}$$

In Eq. (4), $\varepsilon$ denotes the emissivity of the emitting surface, $h$ is Planck's constant, $k_B$ is Boltzmann's constant, and $c$ is the speed of light. Figure 7 shows the spectral intensity as a function of frequency for several HS temperatures assuming maximum emissivity ($\varepsilon = 1$). One of the primary engineering objectives of the HS design is therefore to maximize the emissivity of the emitting surface over the ITER ECE frequency range. The inset of Figure 7 provides a zoomed view over the ITER ECE operational frequency range.

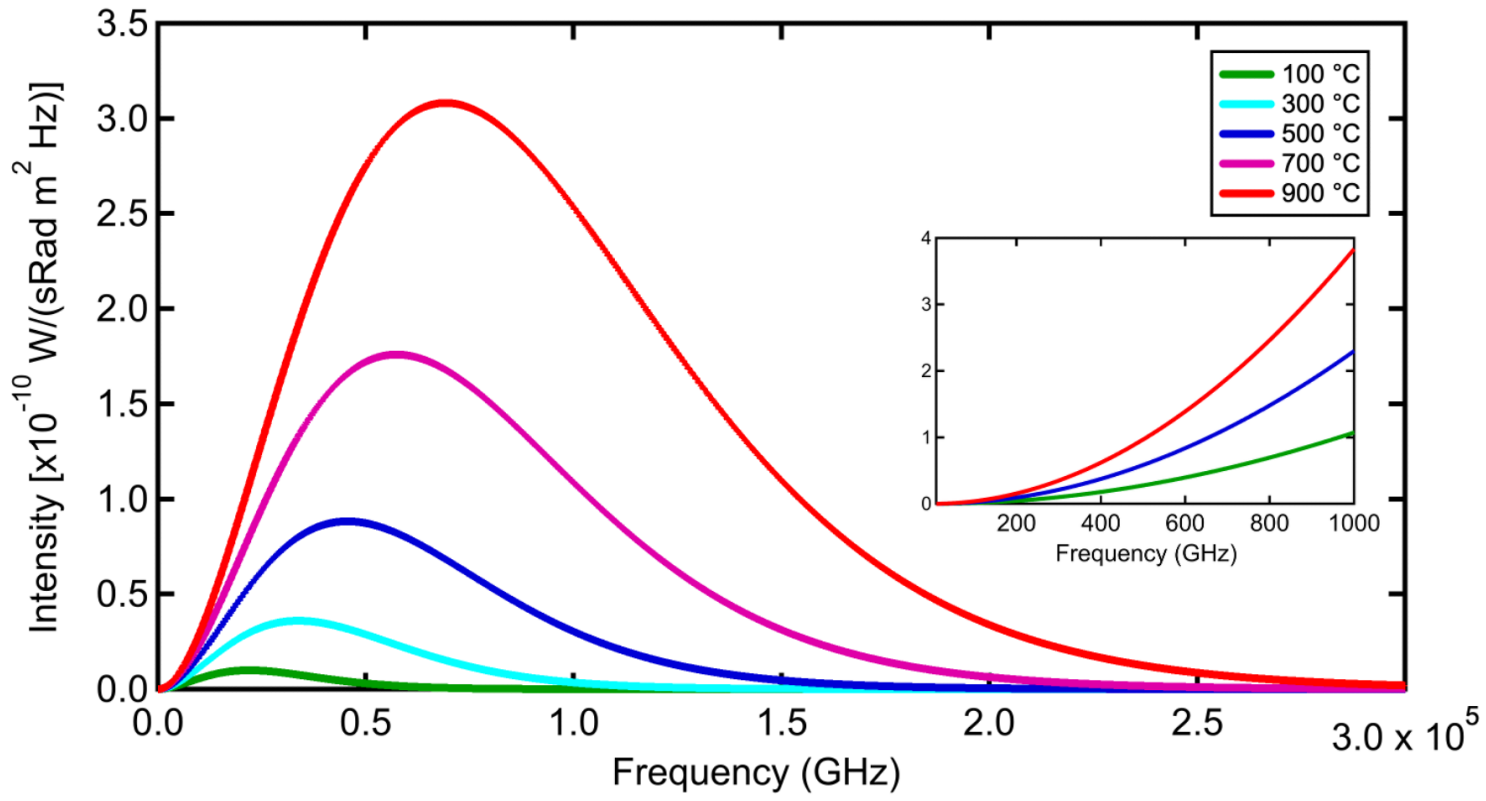

**Figure 7.** Blackbody emission as a function of frequency for different temperatures. The inlet shows the emission for ITER ECE range of frequencies.

In summary, the in-situ hot calibration source enables absolute electron temperature measurements in the ITER ECE diagnostic system. To ensure efficient transmission and calibration performance, the HS emissive surface must satisfy the same $3w$ Gaussian beam criterion used throughout the QO design. However, engineering constraints within DSM2 limit the diameter of the HS emissive surface to 150 mm. As a consequence, the blackbody emission from the HS becomes severely limited below approximately 115 GHz, as shown in Figure 8. The figure indicates that the emitted power at frequencies below 115 GHz is insufficient for reliable calibration of the ECE instruments.

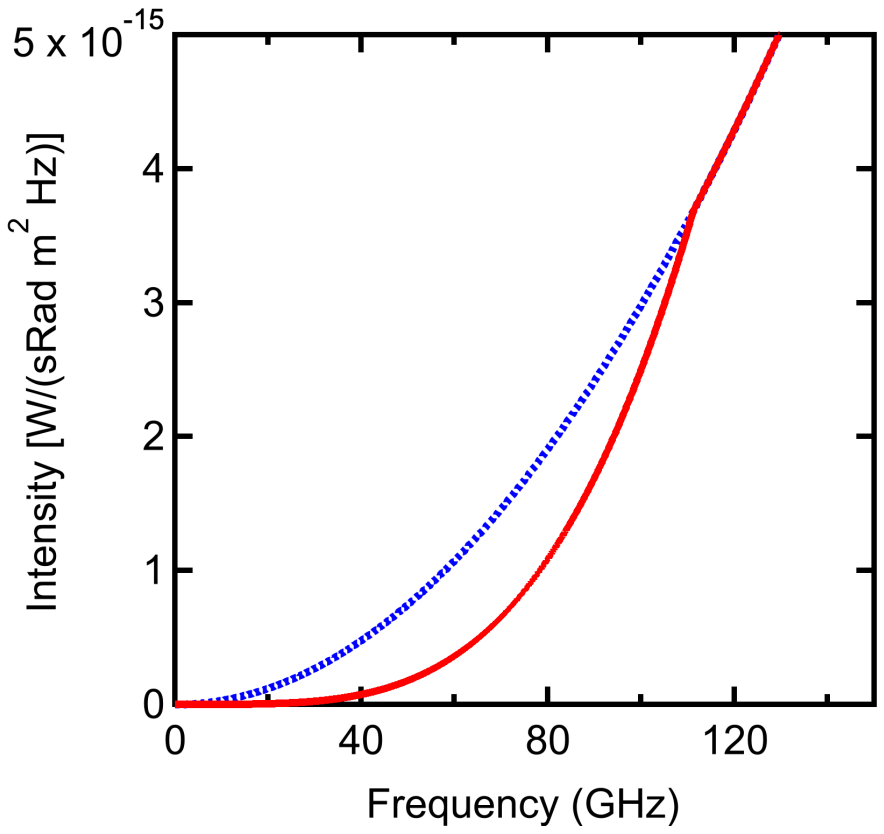

**Figure 8.** Comparison of blackbody emission (dotted curve) from HS as a function of frequency at 700 °C. Finite HS dimensions significantly reduce usable blackbody emission (solid curve) below approximately 115 GHz.

Now that the HS emissive surface size defines the minimum calibrated ECE frequency, this frequency is subsequently used to determine the corresponding $3w$ beam diameter throughout the quasi-optical path. The optical components are then sized such that they accommodate the $3w$ beam envelope at this minimum frequency, thereby maximizing transmitted power across the full ITER ECE frequency range. This sizing criterion is applied not only to the front-end mirrors, but also to the bore diameters in the neutron shielding panels (shown in Figure 1) and the dimensions of the window assembly. In addition, the same $3w$-based approach is used in the design of the PSU optics, including the mirror and the grid polarizer sizes. Figure 9 shows three times the Gaussian beam radius ($3w$) along the optical path for several representative ECE frequencies. Since the beam size is largest at 115 GHz, the $3w$ values corresponding to this frequency are used as the governing design parameters for the optical component dimensions.

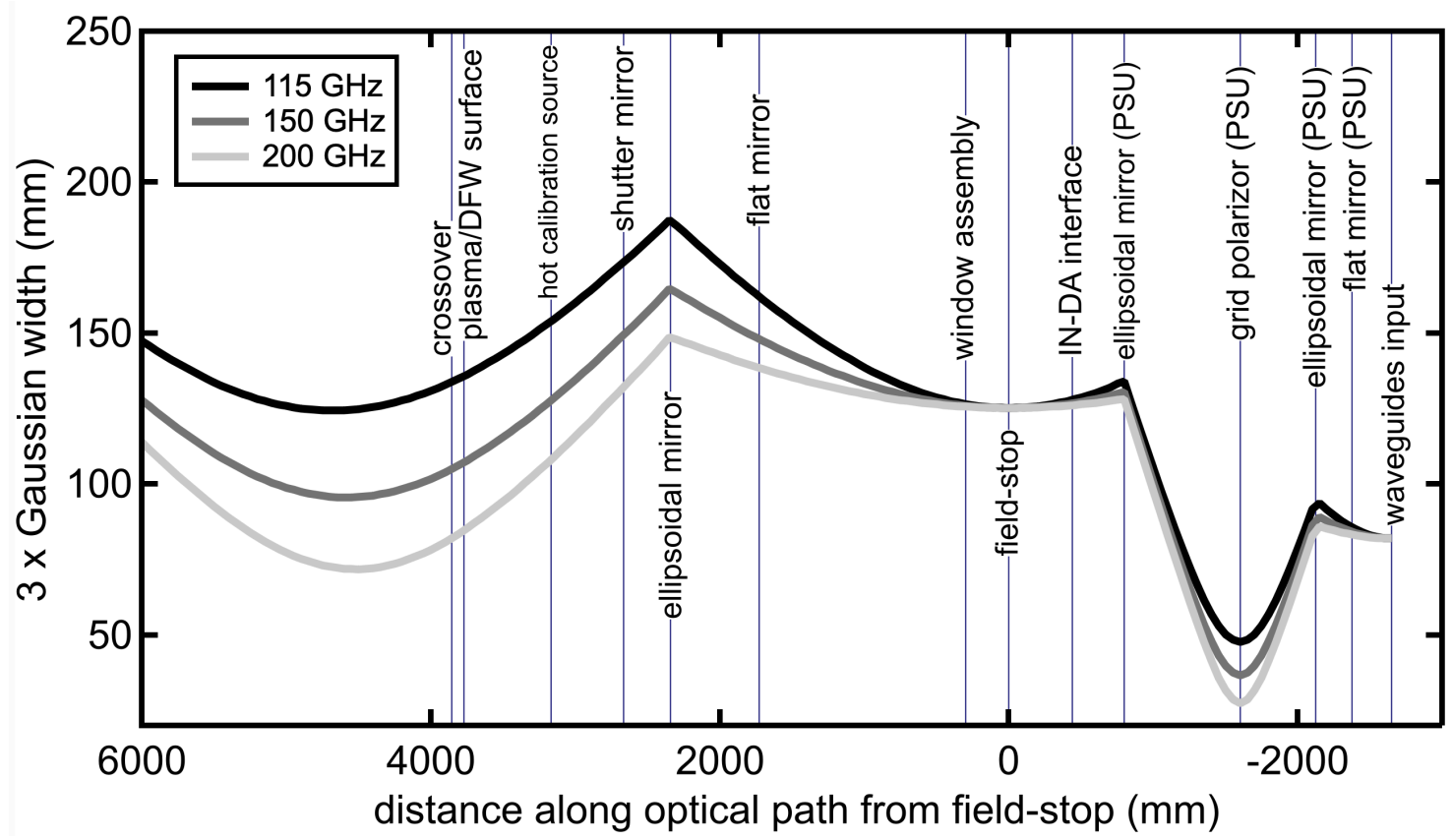

**Figure 9.** Three times the Gaussian beam radius (3w) along the ITER ECE quasi-optical path, referenced to the field-stop located at 0, for representative ECE frequencies.

An independent three-dimensional Gaussian Beam Mode (GBM) analysis was subsequently performed to generate a full 3D rendering of the Gaussian beam propagation using the 3*w* beam envelope. As shown in Figure 1, the 3D 3*w* beam corresponding to 115 GHz was integrated directly into the ITER ECE front-end model. This independent 3D beam analysis confirms the beam variation along the QO path and is in agreement with the 1D Gaussian beam calculations presented earlier, thereby validating the 1D design approach.

The 3D model was further used to verify that the beam propagation remains free of mechanical interference with the DSM2 structure while ensuring that the beam adequately fills the optical components and window assembly to maximize transmission efficiency. Therefore, the 3D results directly informed the DSM design, optical component sizing, and plasma sampling volume estimates.

### 3.2. System misalignment and its consequences

In the current ITER ECE design, the PSU is mechanically fixed with respect to the port cell floor, whereas the front-end assembly, including the field-stop, is integrated into DSM2 as part of the ITER vacuum vessel. Consequently, any displacement of the vacuum vessel results in relative motion between the front-end and the PSU. Such motion may arise from thermal expansion during vessel temperature variations or from electromagnetic loading associated with plasma off-normal events, such as vertical displacement events.

Regardless of the feasibility of potential misalignment mitigation strategies, relative displacement between the front-end and PSU can lead to partial Gaussian beam clipping and, therefore, degradation of the transmitted ECE signal. From a Gaussian beam optics perspective, the misalignment effectively shifts the beam relative to the field-stop, producing an effect equivalent to a reduction in the effective field-stop size.

Figure 10 illustrates the Gaussian beam variation along the quasi-optical path for a 115 GHz beam when the effective field-stop radius is reduced by 10% relative to its nominal value. As shown, the reduced effective field-stop size results in larger beam widths throughout the DSM region. Consequently, maintaining maximum power transmission efficiency under such conditions would require larger mirrors in the front-end as well as larger optical elements within the PSU. More importantly, the increased beam size inside the plasma corresponds to degraded spatial resolution due to the enlarged ECE sampling volume. In addition, the minimum beam-width location shifts toward the plasma edge, indicating that the relatively uniform spatial resolution originally achieved across the plasma minor radius, is no longer maintained.

These results demonstrate that misalignment between the front-end and PSU can significantly degrade ECE diagnostic performance by effectively reducing the field-stop radius and altering the intended Gaussian beam propagation. Therefore, minimizing relative motion between the front-end and PSU is critical for preserving spatial resolution and maintaining reliable ITER ECE measurements.

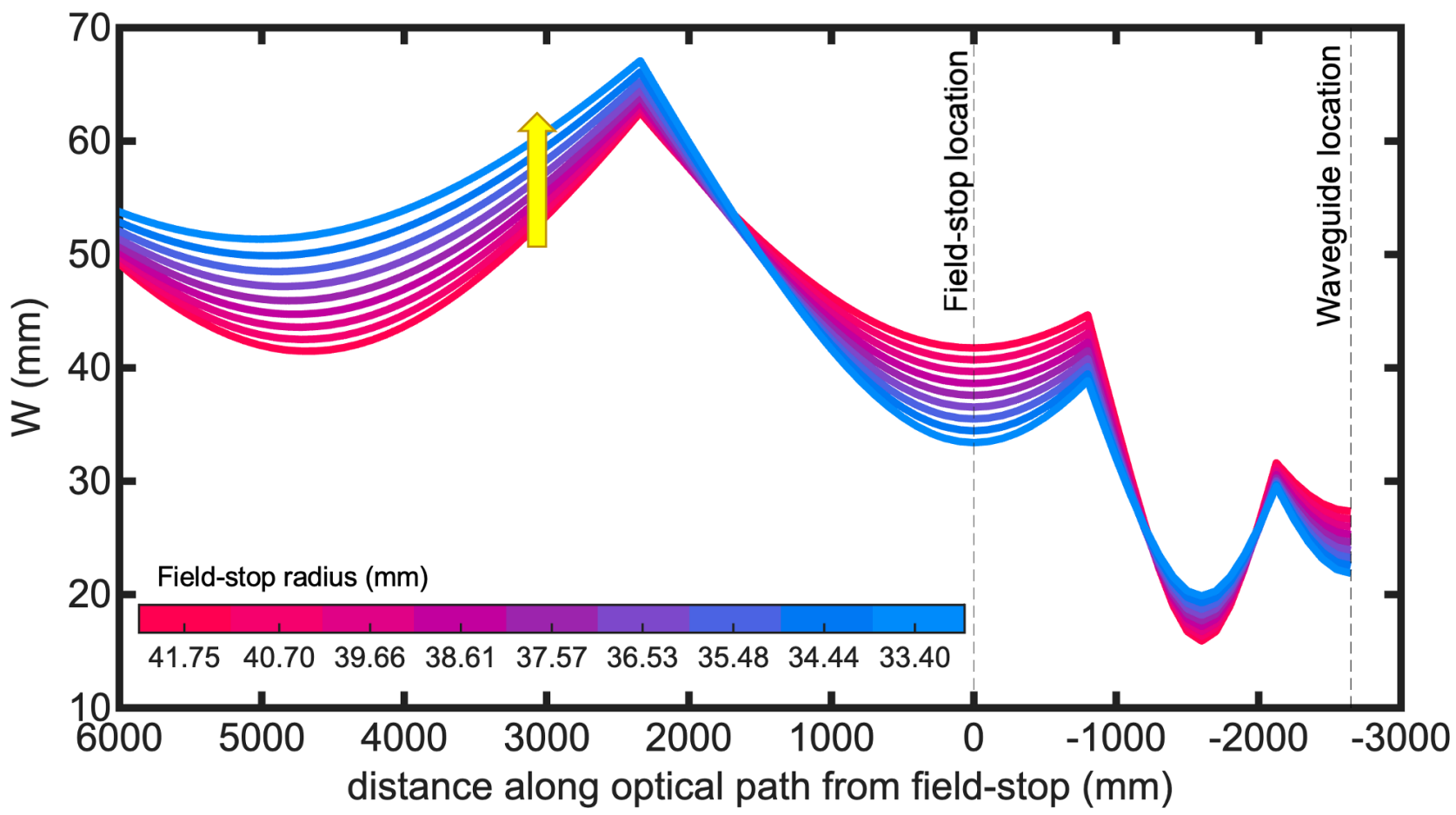

**Figure 10.** Gaussian beam width along ITER ECE quasi-optical path, for beam at 115 GHz, when field-stop radius (located at 0) decreases by 10%. The arrow indicates the increase in beam width within the front-end and plasma regions caused by reduction of the effective field-stop radius.

## 4. Measurements from oblique view

Long-standing and systematic discrepancies between electron temperature measurements obtained from radially viewing ECE diagnostics and Thomson scattering (TS) systems in contemporary tokamaks have motivated renewed interest in the role of non-Maxwellian electron velocity distribution functions (EVDFs) in ECE interpretation. Since ITER is intended not only to support plasma operation and control, but also to advance the scientific understanding required for future reactor-relevant fusion devices, it is essential that the ECE diagnostic provide reliable measurements that can be cross-validated against independent diagnostics such as TS. Resolving the ECE-TS discrepancy is therefore of considerable importance for accurate electron temperature measurements in ITER and future burning plasma experiments.

To address this issue and investigate the impact of non-Maxwellian electron populations on ECE measurements, an oblique viewing geometry was incorporated into the ITER ECE quasi-optical design. The oblique view was initially designed with a viewing angle of 12.5°; however, engineering and spatial constraints within DSM2 ultimately reduced the achievable angle to 9.25°. Despite this reduction, recent studies have demonstrated that the ITER oblique view retains substantial sensitivity to non-thermal electron populations and provides information complementary to the radial view.[11]

Unlike the radial view, the oblique geometry enhances sensitivity to the parallel component of the electron momentum, thereby enabling access to momentum-space information and providing additional insight into the electron distribution function. Recent GENRAY simulations have shown that the oblique view preferentially samples higher-energy electrons and can therefore improve sensitivity to non-thermal features expected in high-temperature ITER plasmas. These simulations utilized kinetic profiles and magnetic equilibrium reconstructed from a recent ITER H-mode scenario at full magnetic field. The analysis also utilized a non-Maxwellian EVDF. The non-Maxwellian EVDF was modeled as an isotropic relativistic two-temperature distribution with modifications spanning $0.75 < u/u_{th} < 1.5$, where momentum $u = \gamma mv$, $u_{th}$ is the electron thermal momentum, and $\gamma$ is the Lorentz factor. The corresponding simulated radiation temperature ($T_{rad}$) spectra for radial and oblique views are shown in Figure 11.*a* and *b*, respectively. These simulations indicate that Doppler broadening becomes the dominant mechanism influencing the oblique ECE spectrum, exceeding both relativistic and non-thermal emissions. Consequently, the oblique view is expected to play an important role in the interpretation of ECE measurements in ITER scenarios where non-thermal electron populations are present.

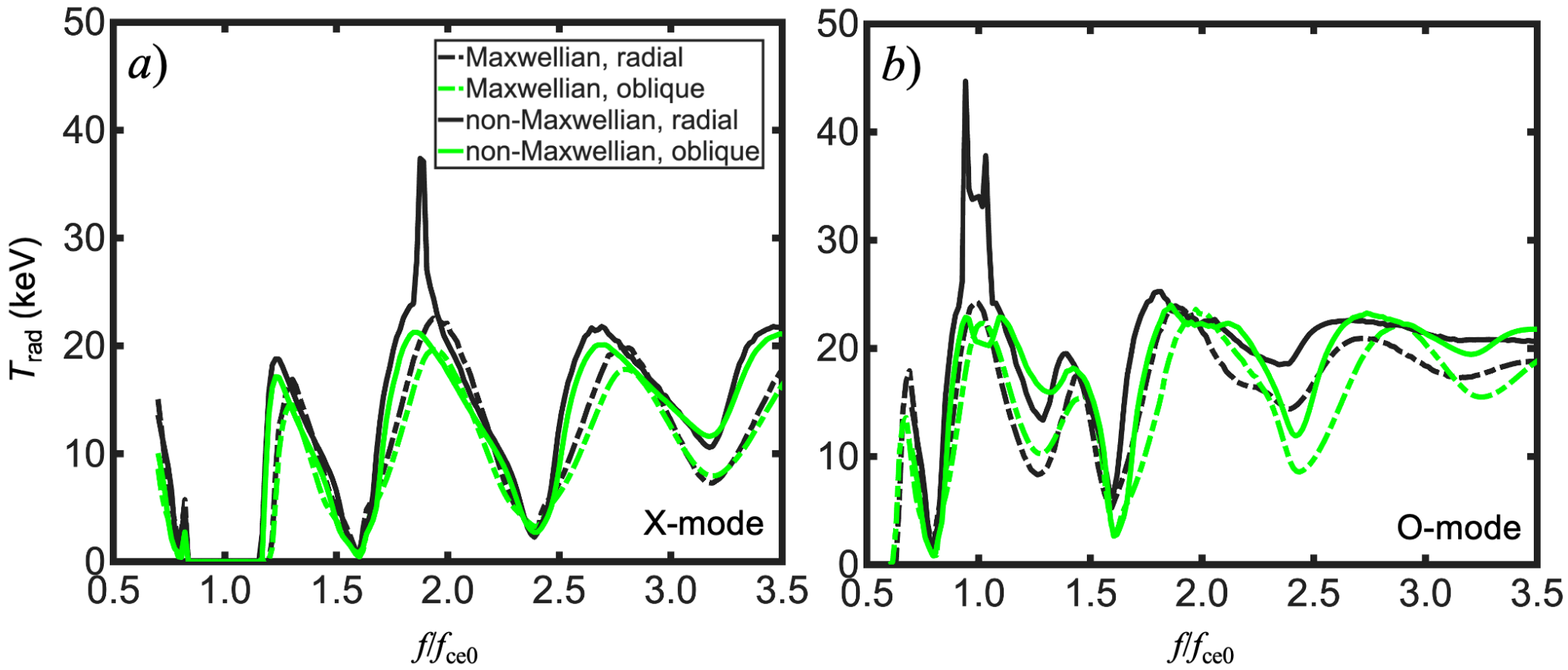

**Figure 11.** Comparison of simulated ECE spectra obtained using Maxwellian and non-Maxwellian electron velocity distribution functions for *a*) X-mode polarization and *b*) O-mode polarization, for both radial and oblique viewing geometries. The simulations were performed using kinetic profiles from ITER discharge 130506, as described in Ref. 11.

Another important mission of the ITER ECE diagnostic is the detection and characterization NTMs. In particular, it is important to assess whether the oblique view can still support NTM detection in scenarios where the radial view becomes compromised. To investigate this, additional ECE simulations were performed. In these simulations, the electron temperature profile from the ITER IMAS scenario shown in Figure 12.*a* was artificially modified to include localized $T_e$-flattening representative of an NTM structure, while the EVDF was assumed to remain Maxwellian.

The simulations demonstrated that the oblique view is capable of diagnosing NTM-induced profile modifications. As shown in Figure 12, the simulated $T_{rad}$ profiles exhibit the same flattening behavior in both radial and oblique views. However, the oblique-view spectra are Doppler shifted in frequency space due to the viewing geometry, and proper interpretation therefore requires corrections based on plasma rotation and parallel refractive index calculations. In addition, the core measurements are affected by changes in the ECE line of sight associated with the oblique geometry. Nevertheless, the results indicate that the oblique view can provide valuable complementary capability for NTM detection and interpretation in ITER.

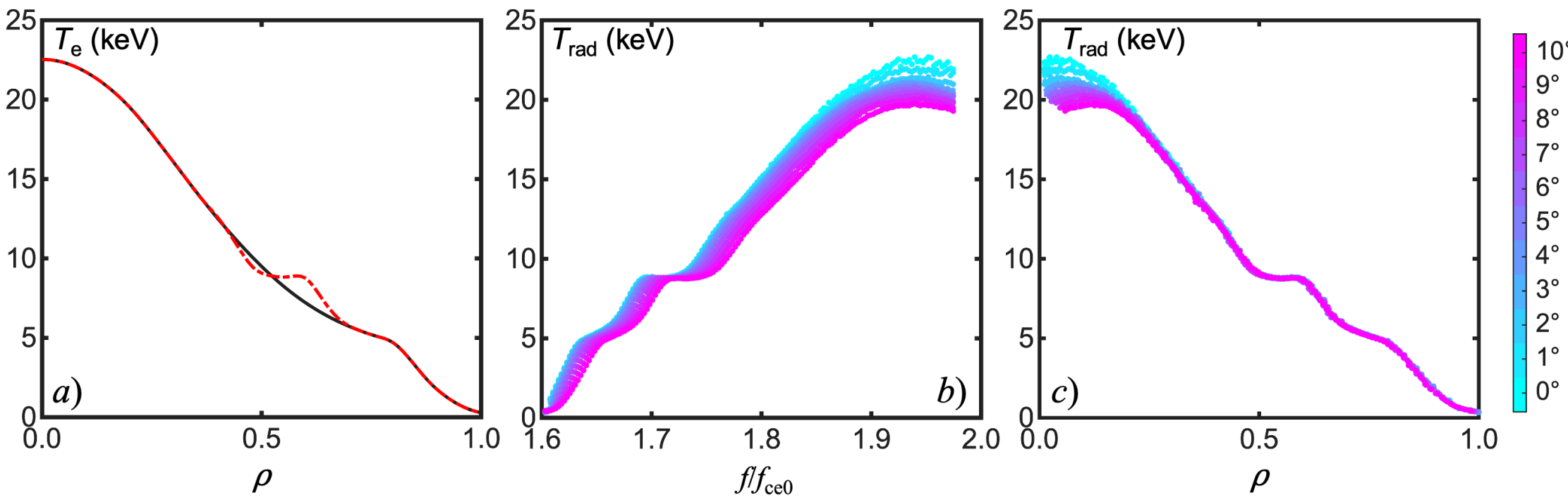

**Figure 12.** *a*) Electron temperature profile versus $\rho$, with artificial flattening representing the effect of a NTM (dashed curve), compared to the original temperature profile (solid curve). *b*) Simulated radiation temperature spectra for X-mode polarization as a function of ECE frequency normalized to the electron cyclotron frequency (cyclotron harmonics). *c*) Simulated $T_{rad}$ profile versus $\rho$ , where $\rho$ is the normalized toroidal flux function. In panels *b* and *c*, the color scale indicates the oblique viewing angles used in the simulations.

## 5. Summary

The quasi-optical (QO) design of the ITER ECE diagnostic system has been presented with emphasis on the role of the field-stop in defining Gaussian beam propagation and overall system performance. The field-stop-based design framework governs the beam evolution throughout the front-end, polarization splitter unit (PSU), and transmission path, while simultaneously determining the plasma sampling volume and spatial resolution of the diagnostic.

A Gaussian beam transmission analysis was performed to evaluate the impact of finite optical apertures and hot calibration source (HS) dimensions on the ITER ECE design. The $3w$ criterion was adopted as the primary design guideline for sizing optical components, including mirrors, bores in the neutron shielding panels, window assemblies, optics in the polarization splitter unit (PSU), and the HS emissive surface. Engineering constraints within DSM2 limit the HS emissive surface diameter to 150 mm, which in turn establishes a minimum calibrated ECE frequency of approximately 115 GHz.

The impact of system misalignment between the front-end and PSU was also investigated. The analysis demonstrated that relative motion between these components effectively reduces the field-stop radius, resulting in increased Gaussian beam width, degraded spatial resolution, and shifts in the plasma sampling volume. These effects highlight the importance of maintaining alignment between the front-end and PSU to preserve diagnostic performance.

Finally, the role of the oblique ECE view in ITER was examined. The oblique geometry was shown to provide enhanced sensitivity to non-thermal electron populations and to extend ECE measurements into momentum-space information relevant to non-Maxwellian electron velocity distribution functions. Simulations further demonstrated that the oblique view can support the diagnosis of neoclassical tearing modes (NTMs), even under conditions where radial measurements may become compromised.

Integrated design and physics understanding are essential for reliable ITER ECE measurements. In ITER, the performance of the ECE diagnostic cannot be separated from the quasi-optical design choices that govern beam propagation, calibration capability, spatial resolution, and viewing geometry. Likewise, accurate interpretation of ECE measurements in high-temperature burning plasmas requires proper consideration of non-thermal electron populations, Doppler effects, and viewing-angle-dependent physics. Consequently, the successful operation of the ITER ECE diagnostic relies on the combined understanding of optical engineering, calibration methodology, and plasma emission physics.

## Acknowledgments

This work is supported PPPL subcontract S013464-C via U.S. DOE Contract No. DE-AC02-09CH11466 with Princeton University.

## Disclaimers

The views and opinions expressed herein do not necessarily reflect those of the ITER organization.

## ORCID IDs

S. Houshmandyar, https://orcid.org/0000-0002-4738-3569
W. L. Rowan, https://orcid.org/0000-0001-8793-919X
J. P. Ziegel, https://orcid.org/0009-0001-2329-8829
A. Ouroua, https://orcid.org/0000-0002-6990-6789